\documentclass[12pt,a4paper]{article}
\usepackage{epsfig}
\begin{document}
\title{Little Higgs models and single top production \\
at the LHC}
\author{Chong-Xing Yue, Li Zhou, Shuo Yang\\
{\small Department of Physics, Liaoning Normal University, Dalian
116029, China
\thanks{E-mail:cxyue@lnnu.edu.cn}}\\}
\date{\today}
\maketitle
\begin{abstract}
\hspace{5mm}We investigate the corrections of the littlest Higgs(LH)
model and the $SU(3)$ simple group  model to single top production
at the CERN Large Hardon Collider(LHC). We find that the new gauge
bosons $W_{H}^{\pm}$ predicted by the LH model can generate
significant contributions to single top production via the s-channel
process. The correction terms for the tree-level $Wqq'$ couplings
coming from the $SU(3)$ simple group model can give large
contributions to the cross sections of the t-channel single top
production process. We expect that the effects of the LH model and
the $SU(3)$ simple group model on single top production can be
detected at the LHC experiments.
\end {abstract}

\vspace{0.8cm}

\newpage
\section*{1. Introduction}

\hspace{5mm}The top quark, with a mass of the order of the
electroweak scale $m_{t}=172.7\pm2.9GeV$[1] is the heaviest particle
yet discovered and might be the first place in which the new physics
effects could be appeared. The properties of the top quark could
reveal information regarding flavor physics, electroweak symmetry
breaking(EWSB) mechanism, as well as new physics beyond the standard
model(SM)[2]. Hadron colliders, such as the Tevatron and the CERN
Large Hadron Collider(LHC), can be seen as top quark factories. One
of the primary goals for the Tevatron and the LHC is to determine
the top quark properties and see whether any hint of non-SM effects
may be visible. Thus, studying the top quark production at hadron
colliders is of great interest. It can help the collider experiments
to probe EWSB mechanism and test the new physics beyond the SM.

In the context of the SM, top quark can be produced singly via
electroweak interactions involving the $Wtb$ vertex. There are
three production processes which are distinguished by the
virtuality $Q^{2}$ of the electroweak gauge boson
$W(Q^{2}=-p^{2}$, where $p$ is the four-momentum of the gauge
boson $W)$[2]. The first process is the so-called $W$-gluon
fusion, or t-channel process, which is the dominant process
involving a space-like $W$ boson($p^{2}<0$) both at the Tevatron
and the LHC. If a $b$ quark distribution function is introduced
into the calculation, the leading order process for the $W$-gluon
fusion channel is the t-channel process $q+b\rightarrow q'+t$
including $\overline{q}'+b\rightarrow \overline{q}+t$[3]. The
second process is the s-channel process
$q+\overline{q}'\rightarrow t+\overline{b}$ mediated by a
time-like $W$ boson ($p^{2}>(m_{t}+m_{b})^{2}$). Single top quark
can also be produced in association with a real $W$
boson($p^{2}\approx M_{W}^{2}$). The cross section for the $tW$
associated production process is negligible at the Tevatron, but
of considerable size at the LHC, where the production cross
section is larger than that of the s-channel process.

At the leading order, the production cross sections for all of three
processes are proportional to the Cabibbo-Kobayashi-Maskawa(CKM)
matrix element $|V_{tb}|^{2}$. Thus, measuring the cross section of
single top production generally provides a direct probe of
$|V_{tb}|$, the effective $Wtb$ vertex, further the strength and
handedness of the top charged-current couplings. This fact has
already motivated large number of dedicated experimental and
theoretical studies. Since the cross section of single top
production is smaller than that of the $t\overline{t}$ production
and the final state signals suffer from large background, the
observation of the single top events is even more challenging than
$t\overline{t}$. It is expected that increased luminosity and
improved methods of analysis will eventually achieve detection of
single top events. So far, there are not single top events to be
observed. The cross sections of single top production  for the s-
and t-channels might be observed at the Tevatron Run II with a small
data sample of only a few $fb^{-1}$. However, the LHC can precisely
measure single top production, the CKM matrix element $V_{tb}$ could
be measured down to less than one percent error at the ATLAS
detector[4].

The three processes for single top production can be affected by new
physics beyond the SM in two ways. One way proceeds via the
modification of the SM couplings between the known particles, such
as $Wtb$ and $Wqq'(q, q'= u, d, c, s)$ couplings, and other way
involves the effects of new particles that couple to the top quark.
Certainly, these two classifications can be seen to overlap in the
limit, in which the extra particles are heavy and decouple from the
low energy description. The SM couplings between the ordinary
particles take well defined and calculable values in the SM, any
deviation from these values would indicate the presence of new
physics. Thus, single top production at hadron colliders might be
sensitive to certain effects of new physics and studying the non-SM
effects on single top production is very interesting and needed.

To address EWSB and the hierarchy problem in the SM, many
alternative new physics models, such as supersymmetry, extra
dimensions, topcolor, and the recently little Higgs models, have
been proposed over the past three decades. Of particular interest to
us is the little Higgs models[5]. In this kind of models, the Higgs
particle is a pseudo-Goldstone boson of a global symmetry which is
spontaneously broken at some high scales. EWSB is induced by
radiative corrections leading to a Coleman-Weinberg type of
potential. Quadratic divergence cancellations of radiative
corrections to the Higgs boson mass are due to contributions from
new particles with the same spin as the SM particles. Some of these
new particles can generate characteristic signatures at the present
and future collider experiments[6,7]. The aim of this paper is to
study the effects of the little Higgs models on single top
production and see whether the corrections of the little Higgs
models to the cross section of single top production can be detected
at the LHC.

The rest of this paper is organized as follows. In the next section,
we shall briefly summarize some coupling expressions in the little
Higgs models, which are related to single top production. The
contributions of the correction terms for the tree-level $Wtb$ and
$Wqq'$ couplings to single top production at the LHC are calculated
in section 3. In section 4, we discuss the corrections of the new
charged gauge bosons, such as $W_{H}^{\pm}$ and $X^{-}$, predicted
by the little Higgs models, to single top production at the LHC. Our
conclusions and discussions are given in section 5.

\section*{2. The relevant couplings}

\hspace{5mm}There are several variations of the little Higgs models,
which differ in the assumed higher symmetry and in the
representations of the scalar multiplets. According the structure of
the extended electroweak gauge group, the little Higgs models can be
generally divided into two classes[6,8]: product group models, in
which the SM $SU(2)_{L}$ is embedded in a product gauge group, and
simple group models, in which it is embedded in a larger simple
group. The littlest Higgs model(LH)[9] and the $SU(3)$ simple group
model[8,10] are the simplest examples of the product group models
and the simple group models, respectively. To predigestion our
calculation, we will discuss single top production at the LHC in the
context of these two simplest models.

The LH model[9] consists of a nonlinear $\sigma$ model with a
global $SU(5)$ symmetry and a locally gauged symmetry
$[SU(2)\times U(1)]^{2}$. The global $SU(5)$ symmetry is broken
down to its subgroup $SO(5)$ at a scale $f\sim\Lambda_{s}/4\pi\sim
TeV$, which results in 14 Goldstone bosons(GB's). Four of these
GB's are eaten by the gauge bosons($W_{H}^{\pm}, Z_{H}, B_{H}$),
resulting from the breaking of $[SU(2)\times U(1)]^{2}$, giving
them mass. The Higgs boson remains as a light pseudo-Goldstone
boson, and other GB's give mass to the SM gauge bosons and form a
Higgs field triplet. The gauge and Yukawa couplings radiative
generate a Higgs potential and trigger EWSB. In the LH model, the
couplings constants of the SM gauge boson $W$ and the new gauge
boson $W_{H}$ to ordinary particles, which are related to our
calculation, can be written as[11]:
\begin{eqnarray}
g_{L}^{Wtb}&=&\frac{ie}{\sqrt{2}S_{W}}[1-\frac{v^{2}}{2f^{2}}(x_{L}^{2}+c^{2}(c^{2}-s^{2}))]
,\hspace{12mm}g_{R}^{Wtb}=0;\\
g_{L}^{Wqq'}&=&\frac{ie}{\sqrt{2}S_{W}}[1-\frac{v^{2}}{2f^{2}}c^{2}(c^{2}-s^{2})],\hspace{23mm}g_{R}^{Wqq'}=0;\\
g_{L}^{W_{H}tb}&=&g_{L}^{W_{H}qq'}=\frac{ie}{\sqrt{2}S_{W}}\frac{c}{s},\hspace{38mm}g_{R}^{W_{H}tb}=g_{R}^{W_{H}qq'}=0.
\end{eqnarray}
Where $\nu\approx246GeV$ is the electroweak scale and
$S_{W}=\sin\theta_{W}$, $\theta_{W}$ is the Weinberg angle.
$c(s=\sqrt{1-c^{2}})$ is the mixing parameter between $SU(2)_{1}$
and $SU(2)_{2}$ gauge bosons and the mixing parameter
$x_{L}=\lambda_{1}^{2}/(\lambda_{1}^{2}+\lambda_{2}^{2})$ comes
from the mixing between the SM top quark $t$ and the vector-like
top quark $T$, in which $\lambda_{1}$ and $\lambda_{2}$ are the
Yukawa coupling parameters. The $SU(2)$ doublet quarks $(q,q')$
represent $(u,d)$ or $(c,s)$. In above equations, we have assumed
$V_{tb}\approx V_{ud}\approx V_{cs}\approx1$.

The $SU(3)$ simple group model[8,10] consists of two $\sigma$
model with a global symmetry $[SU(3)\times U(1)]^{2}$ and a gauge
symmetry $SU(3)\times U(1)_{X}$. The global symmetry is
spontaneously broken down to its subgroup $[SU(2)\times U(1)]^{2}$
by two vacuum condensates $<\Phi_{1,2}>=(0,0,f_{1,2})$, where
$f_{1}\sim f_{2}\sim1TeV$. At the same time, the gauge symmetry is
broken down to the SM gauge group $SU(2)\times U(1)$ and the
global symmetry is broken explicitly down to its diagonal subgroup
$SU(3)\times U(1)$ by the gauge interactions. This breaking
scenario gives rise to an $SU(2)_{L}$ doublet gauge bosons
$(Y^{0},X^{-})$ and a new neutral gauge boson $Z'$. Due to the
gauged $SU(3)$ symmetry in the $SU(3)$ simple group model, all of
the SM fermion representations have to be extended to transform as
fundamental (or antifundamental) representations of $SU(3)$, which
demands the existence of new heavy fermions in all three
generations. The fermion sector of the $SU(3)$ simple group model
can be constructed in two ways: universal and anomaly free, which
might induce the different signatures at the high energy collider
experiments. However, the coupling forms of the gauge bosons $W$
and $X$ to the SM quarks can be unitive written as[6]:
\begin{eqnarray}
g_{L}^{Wtb}&=&\frac{ie}{\sqrt{2}S_{W}}(1-\frac{1}{2}\delta_{t}^{2}),\hspace{24mm}g_{R}^{Wtb}=0;\\
g_{L}^{Wqq'}&=&\frac{ie}{\sqrt{2}S_{W}}(1-\frac{1}{2}\delta_{\nu}^{2}),\hspace{23mm}g_{R}^{Wqq'}=0;\\
g_{L}^{Xtb}&=&\frac{ie}{\sqrt{2}S_{W}}\delta_{t},\hspace{40mm}g_{R}^{Xtb}=0;\\
g_{L}^{Xqq'}&=&\frac{ie}{\sqrt{2}S_{W}}\delta_{\nu},\hspace{38mm}g_{R}^{Xqq'}=0
\end{eqnarray}
with
\begin{equation}
\delta_{t}=\frac{\nu}{\sqrt{2}f}t_{\beta}\frac{x_{\lambda}^{2}-1}{x_{\lambda}^{2}+t_{\beta}^{2}},
\hspace{30mm}\delta_{\nu}=-\frac{\nu}{2f t_{\beta}}.
\end{equation}
Where $f=\sqrt{f_{1}^{2}+f_{2}^{2}}$,
$t_{\beta}=\tan\beta=f_{2}/f_{1}$, and
$x_{\lambda}=\lambda_{1}/\lambda_{2}$.

Using these Feynmen rules, we will estimate the contributions of the
LH model and the $SU(3)$ simple group model to single top production
at the LHC in the following sections.

\section*{3. The contributions of the correction terms
to single \hspace*{1.0cm}top production }

\hspace{5mm}For the t-channel process $q+b\rightarrow q'+t$, at
the leading order, there is only one diagram with $W$ exchange in
the t-channel. In the context of the little Higgs models, the
corresponding scattering amplitude can be written as :
\begin{equation}
M_{i}^{t}=\frac{2\pi\alpha_{e}(1+\delta g_{Li}^{Wtb})(1+\delta
g_{Li}^{Wqq'})}{S_{W}^{2}(\hat{t}-m_{W}^{2})}[\overline{u}(q')\gamma^{\mu}P_{L}u(q)]
[\overline{u}(t)\gamma_{\mu}P_{L}u(b)],
\end{equation}
where $\hat{t}=(P_{b}-P_{t})^{2}$, $P_{L}=(1-\gamma^{5})/2$ is the
left-handed prosection operator. $i=$1 and 2 represent the LH
model and the $SU(3)$ simple group model, respectively. $\delta
g_{Li}^{Wtb}$ and $\delta g_{Li}^{Wqq'}$ are the correction terms
for the $Wtb$ and $Wqq'$ couplings induced by these two little
Higgs models, which have been given in Eqs.(1,2,4,5).

In the context of the little Higgs models, the scattering amplitude
of the s-channel process $q+\overline{q}'\rightarrow t+\overline{b}$
can be written as:
\begin{eqnarray}
M_{i}^{s}=\frac{2\pi\alpha_{e}(1+\delta g_{Li}^{Wtb})(1+\delta
g_{Li}^{Wqq'})}{S_{W}^{2}(\hat{s}-m_{W}^{2})}[\overline{\nu}(\overline{q}')\gamma^{\mu}P_{L}u(q)]
[\overline{u}(t)\gamma_{\mu}P_{L}\nu(\overline{b})],
\end{eqnarray}
where $\hat{s}=(P_{q}+P_{\overline{q}'})^{2}$ and $\sqrt{\hat{s}}$
is the center-of-mass energy of the subprocess
$q+\overline{q}'\rightarrow t+\overline{b}$.

At the leading order, the production of single top quark in
association with a $W$ boson is given via the processes mediated
by the s-channel $b$ quark exchange and the u-channel top quark
exchange. In the little Higgs models, the tree-level coupling of
the gluon to a pair of fermions is same as that in the SM, thus
the scattering amplitude of this process can be written as:
\begin{equation}
M^{tW}_{i}=\frac{eg_{s}(1+\delta
g_{Li}^{Wtb})}{\sqrt{2}S_{W}}\overline{u}(t)[\frac{\not\varepsilon_{2}
P_{L}(\not P_{g}+\not
P_{b}+m_{b})\not\varepsilon_{1}}{\hat{s}'-m_{b}^{2}}+\frac{\not\varepsilon_{1}
(\not P_{t}-\not
P_{g}+m_{t})\not\varepsilon_{2}P_{L}}{\hat{u}-m_{t}^{2}}]u(b),
\end{equation}
where $\hat{s}'=(P_{g}+P_{b})^{2}=(P_{W}+P_{t})^{2}$,
$\hat{u}=(P_{t}-P_{g})^{2}=(P_{b}-P_{W})^{2}$.

After calculating the cross sections $\hat{\sigma_{i}}(\hat{s})$ for
the t-channel, s-channel, and $tW$ associated production processes,
the total cross section $\sigma_{i}(S)$ for each process of single
top production at the LHC can be obtained by convoluting $
\hat{\sigma}_{ijl}(\hat{s})[\hat{\sigma}_{i}(\hat{s})=\sum\limits_{j,l}
\hat{\sigma_{ijl}}(\hat{s})]$ with the parton distribution
functions(PDF's):
\begin{equation}
\sigma_{i}(S)=\sum_{j,l}\int_{0}^{1}dx_{1}\int_{0}^{1}dx_{2}f_{j}(x_{j},\mu_{f}^{2})
f_{l}(x_{l},\mu_{f}^{2})\hat{\sigma_{jl}}(\hat{s}),
\end{equation}
where $j$ and $l$ are the possible combination of incoming gluon,
quark, antiquark. $f(x,\mu_{f}^{2})$ is the PDF evaluated at the
factorization scale $\mu_{f}$. Through out this paper, we neglect
all quark masses with the exception of $m_{t}$, use CTEQ6L PDF with
$\mu_{f}=m_{t}$[12], and take the center-of-mass energy
$\sqrt{S}=14TeV$ for the process $pp\rightarrow t+X$ at the LHC.

To obtain numerical results, we need to specify the relevant SM
parameters. These parameters are $m_{t}$=172.7GeV[1],
$\alpha(m_{Z})$=1/128.8, $\alpha_{s}$=0.118, $S_{W}^{2}$=0.2315, and
$m_{W}$=80.425\\GeV[13]. Except for these SM input parameters, the
contributions of the LH model and the $SU(3)$ simple group model to
single top quark production are dependent on the free parameters
($f$, $x_{L}$, $c$) and ($f$, $x_{\lambda}$, $t_{\beta}$),
respectively. Considering the constraints of the electroweak
precision data on these free parameters, we will assume $f\geq1TeV$,
$0.4\leq x_{L}\leq 0.6$ and $0<c\leq 0.5$ for the LH model[14] and
$f\geq 1TeV$, $x_{\lambda}>1$, and $t_{\beta}>1$ for the $SU(3)$
simple group model[6,8,10], in our numerical estimation.

The relative corrections of the LH model and the $SU(3)$ simple
group model to the cross section $\sigma_{i}$ of single top
production at the LHC  are shown in Fig.1 and Fig.2, respectively.
In these figures, we have taken
$\Delta\sigma_{i}=\sigma_{i}-\sigma^{SM}_{i}$, $f=1.0TeV$ and three
values of the mixing parameters $x_{L}$ and $x_{\lambda}$. From
these figures, we can see that the contributions of the $SU(3)$
simple group model to single top production are larger than those of
the LH model. For the LH model, the absolute values of the relative
correction $\Delta\sigma_{i}/\sigma_{i}^{SM}$ are smaller than $2\%$
in most of the parameter space preferred by the electroweak
precision data. The $SU(3)$ simple group  model has negative
contributions to single top production at the LHC. For $f=1TeV$,
$x_{\lambda}\geq3$, and $1\leq t_{\beta}\leq5$, the absolute values
of the relative correction $\Delta\sigma_{i}/\sigma^{SM}_{i}$ for
the t-channel, s-channel, and $tW$ associated production processes
are in the ranges of $4.3\%\sim10.8\%$, $3.1\%\sim7.5\%$, and
$2\%\sim10.6\%$, respectively.

\begin{figure}[htb] \vspace{-0.5cm}
\begin{center}
\epsfig{file=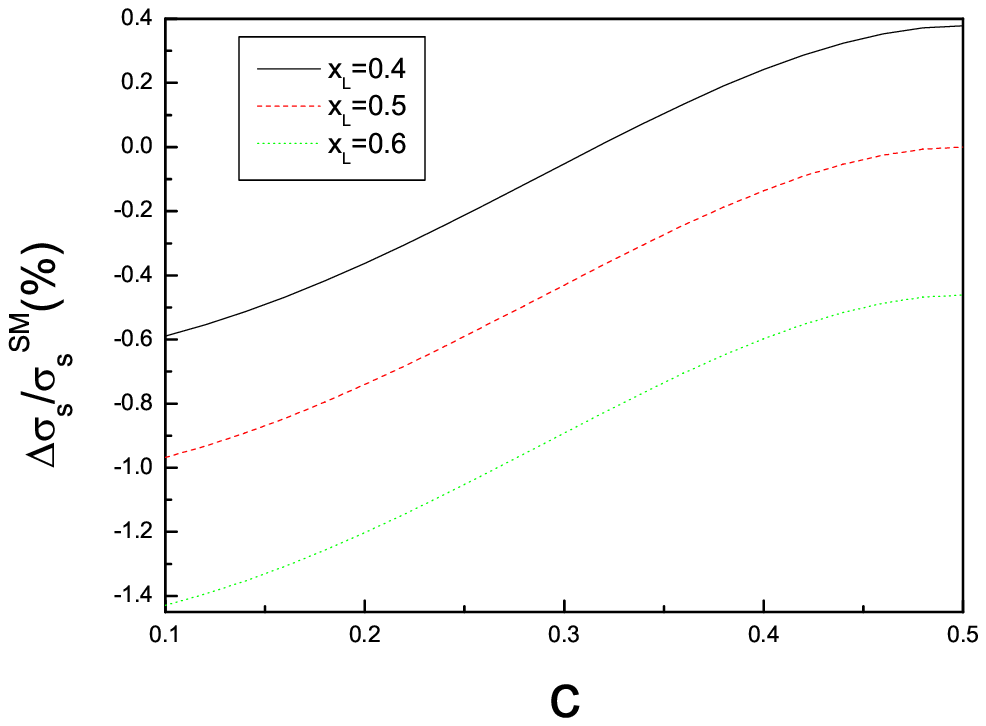,width=220pt,height=205pt}
\hspace{0cm}\vspace{-0.25cm}
\epsfig{file=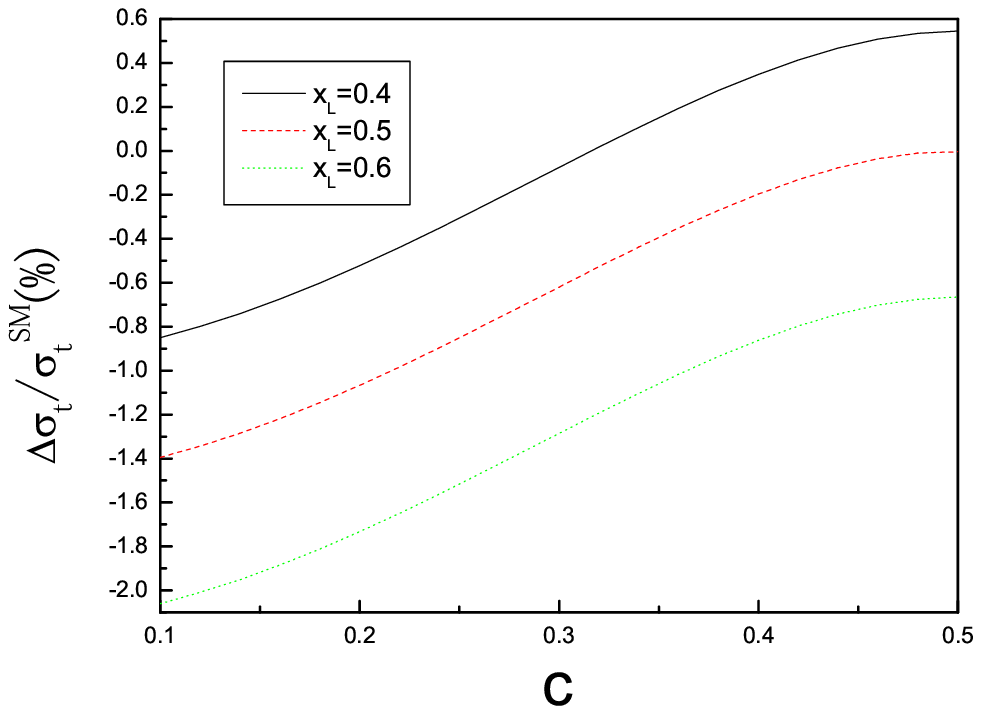,width=220pt,height=207pt} \hspace{-0.5cm}
\hspace{10cm}\vspace{-1cm}
\epsfig{file=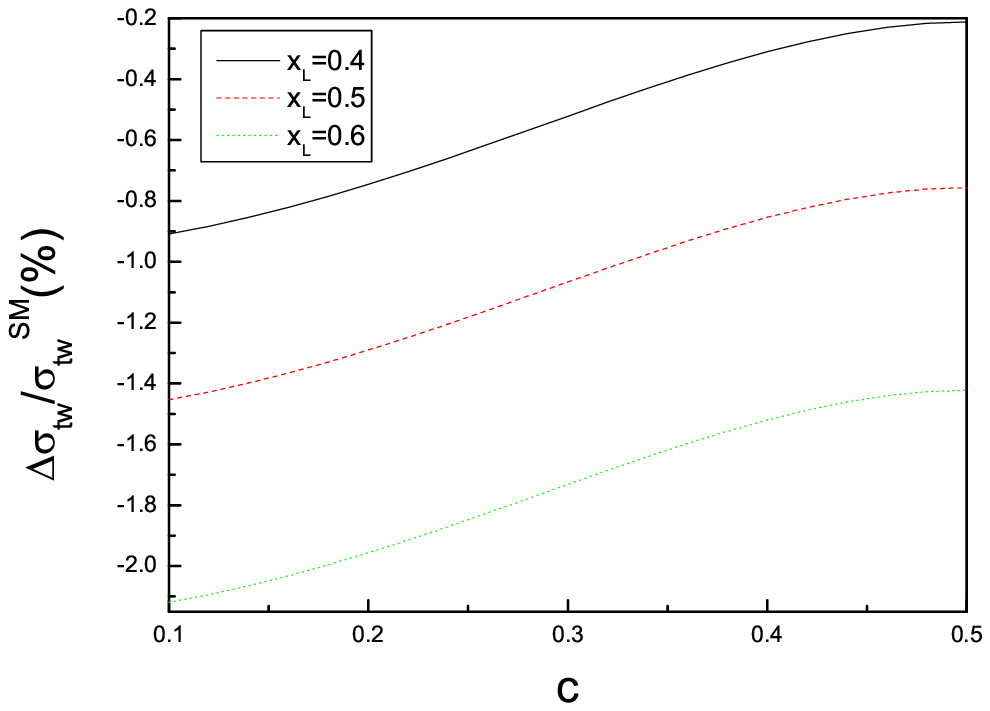,width=220pt,height=200pt} \vspace{+0.2cm}
 \caption{The relative correction $\Delta\sigma_{i}/\sigma^{SM}_{i}$ as a function
 of the mixing parameter $c$ for \hspace*{1.8cm}$f=$1.0TeV and different values of the mixing
parameter $x_{L}$.} \label{ee}
\end{center}
\end{figure}

\begin{figure}[htb] \vspace{-0.5cm}
\begin{center}
\epsfig{file=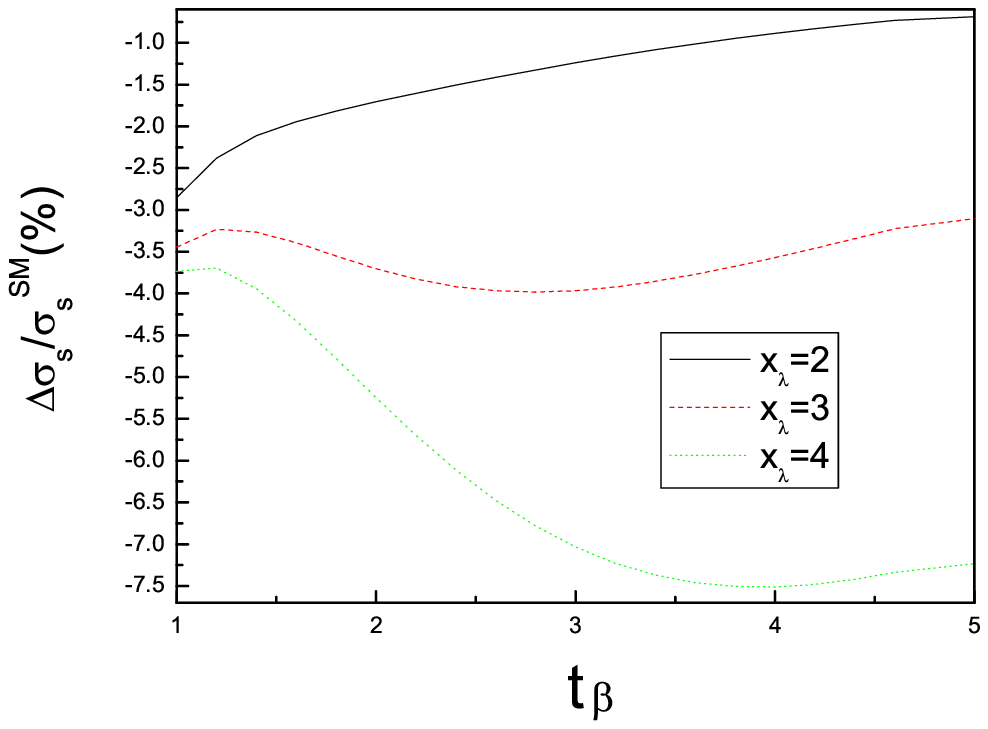,width=220pt,height=205pt}
\hspace{0cm}\vspace{-0.25cm}
\epsfig{file=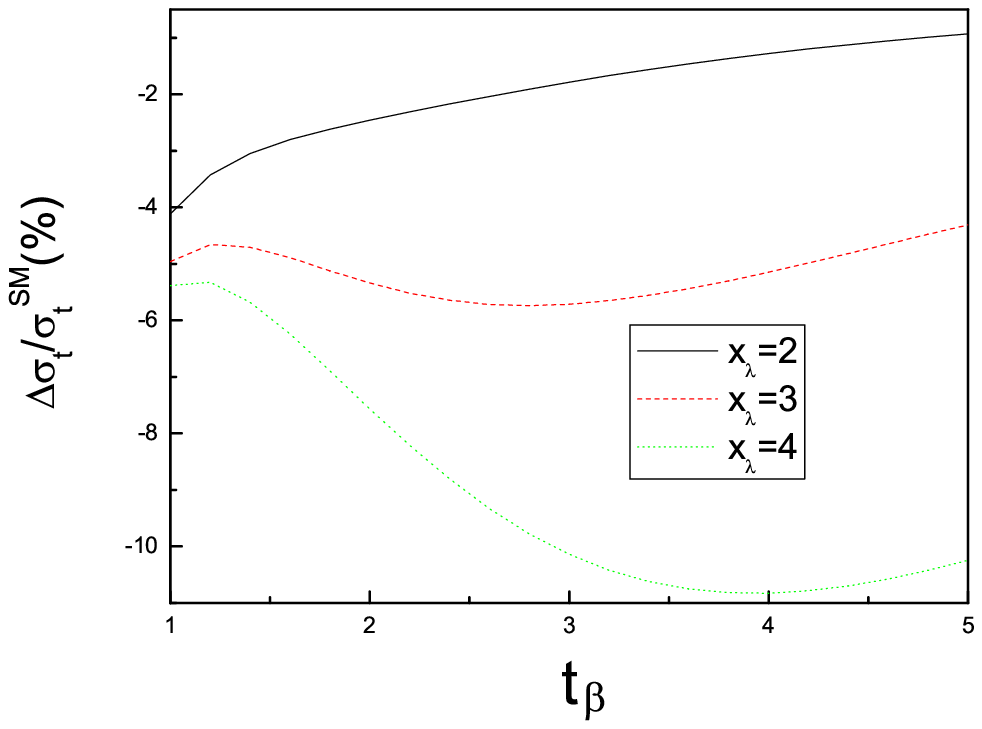,width=220pt,height=207pt} \hspace{-0.5cm}
\hspace{10cm}\vspace{-1cm}
\epsfig{file=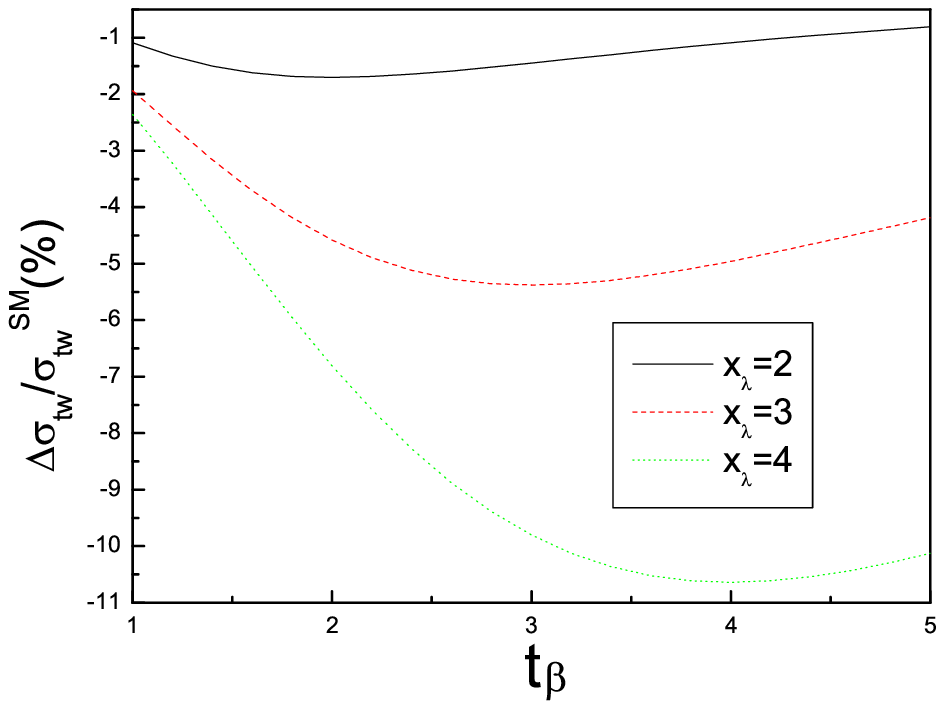,width=220pt,height=200pt} \vspace{0.2cm}
 \caption{The relative correction $\Delta\sigma_{i}/\sigma^{SM}_{i}$ as a function
 of the free parameter
 $t_{\beta}$ for \hspace*{1.9cm}$f=$1.0TeV and different values of the mixing
parameter $x_{\lambda}$.} \label{ee}
\end{center}
\end{figure}

In the context of the SM, the production cross sections at hadron
colliders for the t-channel and s-channel single top production
processes have been calculated at the next leading order(NLO)[15,2].
The values of $\sigma_{t}^{SM}(t)[\sigma_{t}^{SM}(\overline{t})]$
and $\sigma_{s}^{SM}(t)[\sigma_{s}^{SM}(\overline{t})]$ at the LHC
are given as $(156\pm8)pb[(91\pm5)pb]$ and
$(6.6\pm0.6)pb[(4.1\pm0.4)pb]$, respectively. A NLO calculation of
the $tW$ associated production cross section at the LHC has been
recently given in Ref.[16]. The large backgrounds of the signature
from single top production come from W$+$jets and $t\overline{t}$
production. Despite the relatively large expected rate and $D0$ has
developed several advanced multivariate techniques to discriminate
single top production from backgrounds[17], single top production
has not been discovered yet. For all of three processes for single
top production, the production cross section of the t-channel
process can be mostly precise measured at the LHC, which is expected
to be measured to $2\%$ accuracy[4]. Thus, at least we can say that,
in most of the parameter space, the effects of the $SU(3)$ simple
group model on the t-channel single top production process might be
detected at the LHC.

\begin{figure}[htb] \vspace{0.5cm}
\begin{center}
\epsfig{file=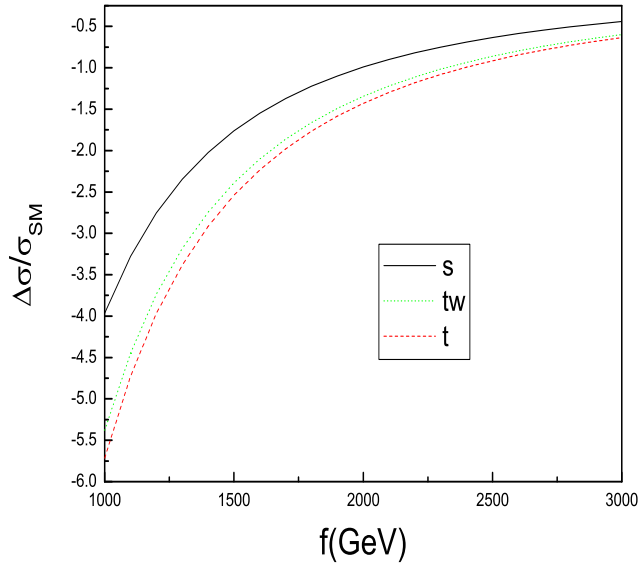,width=270pt,height=250pt} \vspace{-1.0cm}
 \caption{The relative correction $\Delta\sigma_{i}/\sigma^{SM}_{i}$ as a function of
 the scale parameter $f$ for $t_{\beta}$=3 \hspace*{1.8cm}and $x_{\lambda}$=3.}
\label{ee}
\end{center}
\end{figure}

In general, the contributions of the little Higgs models to
observables are proportional to the factor $1/f^{2}$. To see the $f$
dependence of the corrections of the $SU(3)$ simple group model to
single top production, we plot the relative correction
$\Delta\sigma_{i}/\sigma_{i}^{SM}$ as a function of the scale
parameter $f$ for $t_{\beta}=3$ and $x_{\lambda}=3$ in Fig.3, in
which the solid line, dotted line, and dashed line represent the
s-channel, t-channel, and $ tW $ associated production processes,
respectively. One can see from Fig.3 that the value of the relative
correction $\Delta\sigma_{i}/\sigma_{i}^{SM}$ gets close to zero as
$f$ increasing. Thus, the contributions of the $SU(3)$ simple group
model to single top production decouple for large value of the scale
parameter $f$. However, for $t_{\beta}>3$, $x_{\lambda}>3$, and
$1TeV<f\leq2.5TeV$, the absolute value of the relative correction
$\Delta\sigma_{t}/\sigma_{t}^{SM}$ is larger than $2\%$, which might
be detected at the LHC.

\section*{4. The contributions of the new gauge bosons $W_{H}$ and
 \hspace*{0.5cm} $X$ to single top production}

\hspace{5mm}Some of the new particles, such as new charged gauge
boson $W'$ and scalar boson $\Phi$, can couple the top quark to one
of the lighter SM particles and thus can generate contributions to
single top production at tree-level or at one-loop. The one-loop
contributions are generally too small to be observed at hadron
colliders[18]. From Eqs.(3,6,7), we can see that the new charged
gauge bosons $W_{H}^{\pm}$ and $X^{-}$ have contributions to the
t-channel and s-channel processes for single top production.
However, since these new gauge bosons must have space-like momentum
in the t-channel process $q+b\rightarrow q'+t$, their contributions
to the production cross section of the t-channel process are
suppressed by the factor $1/M_{W_{H}}^{2}(M_{X}^{2})$[19]. The
masses $M_{W_{H}}$ and $M_{X}$ are at the order of TeV. Thus, the
contributions of these heavy gauge bosons to the t-channel process
are very small, which can be neglected. For the s-channel process
$q+\overline{q}'\rightarrow t+\overline{b}$, the  new charged gauge
boson $W'$ might generate significant contributions to its
production cross section because of the possibility of $W'$ resonant
production[19,20]. So, in this section, we will only consider that
the contributions of the new gauge bosons $W^{-}_{H}$ and $X^{-}$ to
the s-channel process $\overline{q}+q'\rightarrow \overline{t}+b$.
Certainly, our numerical results are easily transferred to those of
the new charged gauge boson $W^{+}_{H}$ for the process
$q+\overline{q}'\rightarrow t+\overline{b}$ by replacing
$\overline{b}$ as $b$ and $t$ as $\overline{t}$.

The center-of-mass energy $\sqrt{S}$ of the LHC is large enough to
produce the heavy gauge bosons $W_{H}^{-}$ or $X^{-}$ on shell, thus
these heavy gauge bosons might produce significant contributions to
the s-channel process $\overline{q}+q'\rightarrow \overline{t}+b$.
The corresponding scattering amplitude including the SM gauge boson
$W$ can be written as:
\begin{equation}
M_{i}=\frac{2\pi\alpha_{e}}{S_{W}^{2}}[\frac{1}{\hat{s}-m_{W}^{2}}+
\frac{AB}{\hat{s}-M_{i}^{2}+iM_{i}\Gamma_{i}}]
[\overline{u}(\overline{q})\gamma^{\mu}P_{L}\nu(q')]
[\overline{u}(b)\gamma_{\mu}P_{L}\nu(\overline{t})],
\end{equation}
where $i$ represents the gauge boson $W_{H}^{-}$ or $X^{-}$. For the
gauge boson $W_{H}^{-}$, $A=B=c/s$, and for the gauge boson $X^{-}$,
$A=\delta_{t}$ and $B=\delta_{\nu}$. The expression of the total
decay width $\Gamma_{W_{H}}$ has been given in Ref.[21]. If the
decay of the gauge boson $X^{-}$ to one SM fermion and one TeV-scale
fermion partner is kinematically forbidden, then it can decay to
pairs of SM fermions through their mixing with the TeV-scale fermion
partners, which is independent of the fermion embedding[6]. For the
gauge boson $X^{-}$, the possible decay modes are $\overline{t}b$,
$\overline{u}d$, $\overline{c}s$, and $l\nu_{l}$, in which $l$
presents all three generation leptons $e$, $\mu$, and $\tau$. The
total decay width $\Gamma_{X^{-}}$ can be written as[6]:
\begin{equation}
\Gamma_{X^{-}}=\frac{\alpha
M_{X}}{4S_{W}^{2}}(\delta_{t}^{2}+5\delta_{\nu}^{2}).
\end{equation}

The new gauge bosons predicted by the little Higgs models get their
masses from the $f$ condensate, which breaks the extended gauge
symmetry. At the leading order, the masses of the new charged gauge
bosons $W_{H}^{\pm}$ and $X^{-}$ can be written as[5,6,7]:
\begin{eqnarray}
M_{W_{H}}&=&\frac{gf}{2sc}\approx 0.65f\cdot\frac{c}{s},\\
M_{X}&=&\frac{gf}{\sqrt{2}}\approx 0.46f.
\end{eqnarray}

\begin{figure}[htb] \vspace{-1cm}
\begin{center}
\epsfig{file=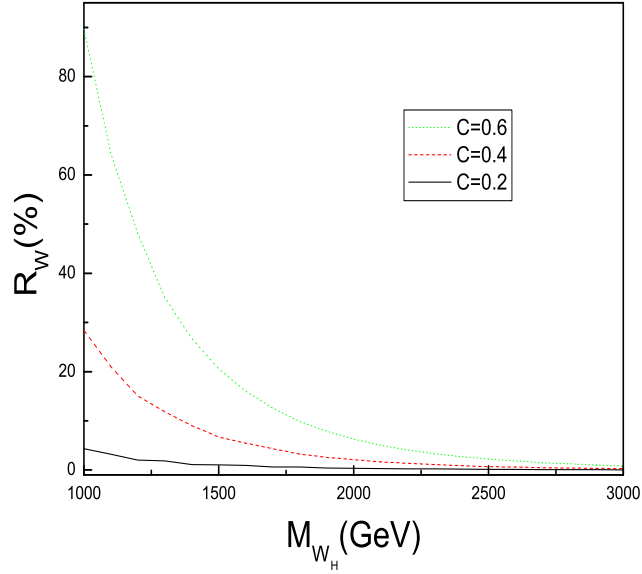,width=270pt,height=250pt} \vspace{-1.0cm}
 \caption{The relative correction parameter $R_{W}$ as a function of
 $M_{W_{H}}$ for three values of  \hspace*{2.0cm}the mixing parameter $c$.}
\label{ee}
\end{center}
\end{figure}

For the LH model, if we assume that the free parameters $f$ and $c$
are in the ranges of $1\sim3TeV$ and $0\sim0.5$, then we have
$M_{W_{H}}\geq1.12TeV$. While the mass of the gauge boson $X^{-}$
predicted by the $SU(3)$ simple group model is larger than $920GeV$
even for $f\geq2TeV$. As numerical estimation, we will simple assume
$1TeV\leq M_{W_{H}}\leq3TeV$ and $1TeV\leq M_{X}\leq3TeV$.

\begin{figure}[htb]
\vspace{-0.5cm}
\begin{center}
\epsfig{file=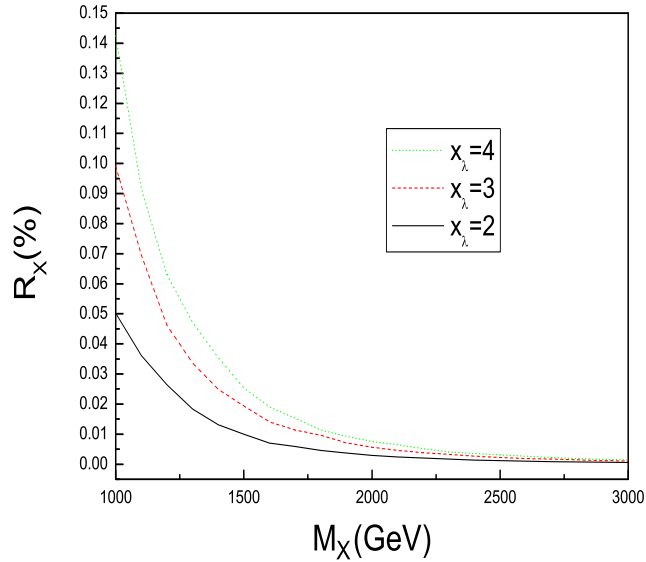,width=270pt,height=250pt} \vspace{-1.0cm}
 \caption{The relative correction parameter $R_{X}$ as a function of
 $M_{X}$ for $t_{\beta}=3$ and \hspace*{2.0cm}three values of the mixing parameter $x_{\lambda}$.}
\label{ee}
\end{center}
\end{figure}

In Fig.4 and Fig.5, we plot the relative correction parameters
$R_{W}=\Delta\sigma_{s}^{W}/\sigma_{s}^{SM}$ and
$R_{X}=\Delta\sigma_{s}^{X}/\sigma_{s}^{SM}$ as functions of the
gauge boson masses $M_{W_{H}}$ and $M_{X}$ for $t_{\beta}=3$ and
three values of the mixing parameters $c$ and $x_{\lambda}$,
respectively. In these figures, we have assumed
$\Delta\sigma_{s}^{W}=\sigma_{s}(W+W_{H})-\sigma_{s}(W)$ and
$\Delta\sigma_{s}^{X}=\sigma_{s}(W+X)-\sigma_{s}(W)$. From these
figures, we can see that the value of $R_{W}$ is significantly
larger than that of $R_{X}$. This is because, compared to the gauge
boson $W_{H}^{-}$, the contributions of the gauge boson $X^{-}$ to
the s-channel process $\overline{q}+q'\rightarrow \overline{t}+b$
are suppressed by the factor $\nu^{2}/f^{2}$. For $0.3\leq c\leq
0.6$ and $1TeV \leq M_{W_{H}}\leq 2TeV$, the value of the relative
correction parameter $R_{W}$ is in the range of $ 1.5\%\leq
R_{W}\leq 90\% $. Even for $M_{W_{H}}\geq2.0TeV(f \sim 2TeV)$, the
value of the relative correction parameter $R_{W}$ can reach
$6.3\%$. Thus, the effects of the new gauge boson $W_{H}$ to the
s-channel process for single top production might be detected at the
LHC.

\section*{5. Conclusions and discussions}

\hspace{5mm}The electroweak production of single top quark at hadron
colliders is an important prediction of the SM which proceeds
through three distinct subprocesses. These  subprocesses are
classified by the virtuality of the electroweak gauge boson $W$
involved: t-channel($p^{2}<0$), s-channel($p^{2}>0$), and associated
$tW$($p^{2}=m_{W}^{2}$) production. Each process has rather distinct
event kinematics, and thus are potentially observable separately
from each other[2]. All of these processes are sensitive to
modification of the $Wtb$ coupling and the s-channel process is
rather sensitive to some heavy charged particles. Thus, studying
single top production at the LHC can help to text the SM and further
to probe new physics beyond the SM.

To solve the so-called hierarchy or fine-tuning problem of the SM,
the little Higgs theory was proposed as a kind of models to EWSB
accomplished by a naturally light Higgs boson. All of the little
Higgs models predict the existence of the new heavy gauge bosons and
generate corrections to the SM tree-level $Wqq'$ couplings. Thus,
the little Higgs models have effects on single top production at
hadron colliders.

Little Higgs models can be generally divided in two classes: product
group models and simple group models. The LH model and the $SU(3)$
simple group model are the simplest examples of the two class
models, respectively. In this paper, we have investigated single top
production at the LHC in the context of the LH model and the $SU(3)$
simple group model. We find that these two simplest little Higgs
models generate contributions to single top production at hadron
colliders via two ways: correcting the SM tree-level $Wqq'$
couplings and new charged gauge boson exchange. For the LH model,
the contributions mainly come from the s-channel $W_{H}$ exchange.
For $0.3\leq c\leq 0.6$ and $1TeV \leq M_{W_{H}}\leq 2TeV$, the
relative correction of the new gauge boson $W_{H}$ to the production
cross section of the s-channel process at the LHC with
$\sqrt{S}$=14TeV is in the range of $ 1.5\%\leq R_{W}\leq 90\% $.
For the $SU(3)$ simple group model, the contributions of the new
gauge boson $X$ to the s-channel single top production process is
very small. However, in most of the parameter space, the correction
terms to the tree-level $Wtb$ and $Wqq'$ couplings can generate
significant corrections to all production cross sections of the
three processes for single top production at the LHC.

It is well known that precise electroweak data provide strong
constraints on any extensions of the SM. Most of the little Higgs
models are severed constrained by the precise electroweak data, with
the exception of the littlest Higgs model with T parity, in which a
low scale parameter $f$ is allowed. However variations in the model
can give rise to very different constraints. For example, for the LH
model, if the SM fermions are charged under $U(1)_{1}\times
U(1)_{2}$, the constraints become relaxed. The scale parameter
$f=1\sim2TeV$ is allowed for the mixing parameter $c$, $c'$, and
$x_{L}$ in the ranges of $0\sim0.5$, $0.62\sim0.73$, and
$0.3\sim0.6$, respectively[6,14]. In this case, the mass of the new
charged gauge boson $W_{H}$ is allowed in the range of
$1TeV\sim3TeV$. Thus, as numerical estimation, we have simply
assumed the scale parameter $f\geq1TeV$. Certainly, the effects of
the little Higgs models on single top production decrease as $f$
increasing, as shown in Fig.3. However, our numerical results shown
that, even for $f\geq2TeV$, the relative correction of the $SU(3)$
simple group model to the cross section for the t-channel single top
production process can reach $-9\%$. Even we assume that the mass of
the new charged gauge boson $W_{H}$ predicted by the LH model is
larger than $2TeV$, it can also make the cross section of the
s-channel single top production process enhance about $6\%$. So we
expect that the effects of the $SU(3)$ simple group model on the
t-channel process for single top production and the contributions of
the new charged gauge bosons $W_{H}^{\pm}$ predicted the LH model to
the s-channel process can be detected at the LHC experiments.

\vspace{0.5cm} \noindent{\bf Acknowledgments}

This work was supported in part by Program for New Century
Excellent Talents in University(NCET-04-0290), the National
Natural Science Foundation of China under the Grants No.10475037.

\null

\end{document}